\documentclass[aps,twocolumn,groupedaddress,showpacs]{revtex4}

\usepackage{graphicx}
\usepackage{epsfig}

\begin{document}

\title{Detecting charge noise with a Josephson junction:\break A problem
of thermal escape in presence of non-Gaussian fluctuations}

\author{Joachim~Ankerhold}

\affiliation{Physikalisches Institut,
Albert-Ludwigs-Universit\"at,
 79104 Freiburg, Germany}

\date{\today}

\begin{abstract}
Motivated by several experimental activities to detect
charge noise produced by a mesoscopic conductor with a
Josephson junction as on-chip detector, the switching rate
out of its zero-voltage state is studied. This process is
related to the problem of thermal escape in presence of
non-Gaussian fluctuations. In the relevant case of weak
higher than second order cumulants, an effective
Fokker-Planck equation is derived, which is then used to
obtain an explicit expression for the escape rate.
Specific results for the rate asymmetry due to the third
moment of current noise allow to analyse experimental data
and to optimize detection circuits.

\end{abstract}


\maketitle

A complete understanding of electronic transport through
mesoscopic conductors necessitates the knowledge of all
noise properties of the corresponding current. This is the
goal of full counting statistics, which has attracted
considerable activities in the last years \cite{lll,bb}.
Theoretically, generating functionals of current noise in
the low frequency limit have been calculated for a variety
of systems comprising tunnel contacts, diffusive wires, and
ballistic cavities \cite{bb}. Experimentally, in a
pioneering measurement the third moment of current noise
produced in a tunnel junction  was detected by analog
amplifiers and filtering techniques in \cite{rsp} and
later also in \cite{bomze}. Since then strong efforts have
been made towards on-chip detection schemes, first because
they are faster and second because they give access to
finite frequency noise properties. Lately, the
distribution of charges flowing through a quantum dot has
been extracted on-chip in the low frequency regime
\cite{ensslin}. The goal now is to push devices
 into higher frequency ranges (GHz), where quantum
effects, electron-electron interactions, and plasmon
dynamics are relevant.

Based on an idea proposed in \cite{tn} and studied later
in various scenarios \cite{heikila,brosco}, currently,
several experiments are aiming to set up circuits with
Josephson junctions (JJs) as detection elements
\cite{lds,pekola1,saclay}. JJs can be fabricated and
manipulated in a very controlled way, offer a large
bandwidth depending on their plasma frequency, and contain
an intrinsic amplification mechanism. Namely, the switching
out of the zero voltage state is exponentially sensitive
to variations of the barrier potential and to the noise
strength. For electrical noise large positive and negative
fluctuations from the mean occur with different
probabilities. The idea is thus to probe this asymmetry by
measuring switching rates for mean mesoscopic currents
flowing forward and backward, respectively. Indeed, recent
experimental results \cite{saclay}
 indicate a sufficient sensitivity of such a circuit to retrieve the third
 cumulant.

A typical set-up \cite{pekola1} consists of a JJ, on which
two currents $I_b$ and $I_m$ are injected. Current $I_b$
is a standard bias current coming from a source in
parallel to the JJ, while $I_m$ runs through a noise
generating mesoscopic conductor in series with the JJ in
such a way that no dc component of $I_m$ passes through
the detector. Due to substantial heating from the
additional electrical noise the JJ operates in the regime
of classical escape, where quantum effects are negligible.
Then, according to the resistively and capacitively
shunted junction model,
 the phase $\varphi$  of the JJ moves in a tilted washboard potential $-E_J
\cos(\varphi)-(\hbar/2e) \langle I_b\rangle \varphi$ with
Josephson energy $E_J$
  and is subject to Johnson-Nyquist noise
$\delta I_b=I_b-\langle I_b\rangle$ and stationary
non-Gaussian current fluctuations $\delta I_m=I_m-\langle
I_m\rangle$. If the phase is initially trapped in one of
the wells (zero-voltage state), it may for sufficiently
large $\langle I_b\rangle <I_c=(2e/\hbar) E_J$ escape so
that the JJ switches to a finite voltage state. Further,
since the third cumulant vanishes in equilibrium due to
time-reversal symmetry, experimentally, the mesoscopic
conductor is at low temperatures driven far from
equilibrium into the shot noise regime, where no
fluctuation-dissipation theorem applies. Hence, the
switching of the JJ can be visualized as the diffusive
dynamics of a fictitious particle in a metastable well
with non-Gaussian continuous fluctuations acting as {\em
external} random driving force.

In the past JJs have been used to thoroughly confirm
various theories for thermal escape
\cite{devoret1,revmodphys} including cases of external
driving either by time-periodic forces or Gaussian noise
sources related to phenomena such as resonant activation
\cite{resoact,flucbarr} and stochastic resonance
\cite{stochres}. When putting the experimental situation
described above into this context, one arrives at a new
type of rate problem: Thermal escape driven by non-Gaussian
continuous noise. Theoretically, the challenge here is
that externally driven escape necessitates a full
dynamical description because the stationary well state
from which particles are ejected over the barrier is only
known {\em a posteriori}. In contrast,  in the undriven
situation thermodynamic approaches, as e.g.\ the bounce
method, can rely on a thermal equilibrium state inside the
well region. In the classical domain and for Gaussian
noise processes a dynamical formulation  is based on a
Fokker-Planck equation (FPE) for the phase space
distribution as shown in a variety of applications
\cite{revmodphys,risken}. Thus, the first goal  is to
derive a generalized FPE for weak non-Gaussian noise,
which then serves as a basis for the rate calculation. A
solution to this problem is of general interest for rate
processes in complex media and crucial for present
electrical noise measurements because non-Gaussian
components are typically very small compared to a
prevailing Gaussian background. In this Letter we develop
the framework for such a rate theory and  give explicit
expressions in case of a JJ as detector for electrical
noise.

For this purpose let us consider the Langevin equation for
the  diffusive motion of a particle of mass $m$
\begin{equation}
m\ddot{\varphi}(t)+ V'(\varphi)-\eta(t)+m\gamma\,
\dot{\varphi}(t)=\xi(t)\ , \label{lange}
\end{equation}
where $\varphi$ denotes a generalized coordinate,
$\dot{}=d/dt$, and $\ '=d/d\varphi$. The barrier potential
$V(\varphi)$ is assumed to be sufficiently smooth with a
well located at $\varphi=0$ with frequency
$\omega_0=\sqrt{V''(0)/m}$ and a barrier top at
$\varphi=\varphi_b$ with frequency
$\omega_b=\sqrt{|V''(\varphi_b)|/m}$ and height
$V_b=V(\varphi_b)-V(0)$. The white thermal Gaussian noise
obeys $\langle \xi(t)\rangle_\beta=0$ and is related to the
damping $\gamma$ via the dissipation fluctuation theorem:
$\langle \xi(t)\xi(t')\rangle_\beta=(2m \gamma/\beta) \,
\delta(t-t')$ with inverse temperature $\beta=1/k_{\rm
B}T$. The statistical properties of the stationary
non-Gaussian noise are determined by the generating
functional
\begin{equation}
{\rm e}^{- S[w(s)]} = \Big\langle {\cal
T}\exp\left[{i}\int_0^t ds \eta(s) w(s)\right]\Big\rangle\,
,
\end{equation}
where ${\cal T}$ is the time ordering operator and the
cumulants are gained from the functional derivatives of
$S[w]$. In particular, we assume $S_1(t)=\langle
\eta(t)\rangle= i\left.\partial S[w]/\partial
w(t)\right|_{w=0}=0$, while the auto-correlation function
reads
\[
S_2(t)=\langle \eta(t)\, \eta(0)\rangle=\left.\partial^2
S[w]/\partial w(t)\partial w(0)\right|_{w=0}\,
\]
and the third cumulant follows accordingly as
\[
S_3(t,t')=\langle \eta(t+t')\, \eta(t')\eta(0)\rangle\, .
\]

Now, particles initially confined in the well region may
escape such that for sufficiently high barriers ($\beta
V_b\gg 1$) a stationary flux appears related to an escape
rate $\Gamma=\langle v\,
\delta(\varphi-\varphi_b)\rangle_{\rm flux}/N_{\rm well}$.
The denominator is the population in the well and the
expectation value of $v=\dot{\varphi}$ is taken with
respect to a quasi-stationary nonequilibrium state, the
flux state, which may be cast into the form
\begin{equation}
P_{\rm flux}(v ,\varphi)=P_{\rm equi} (v,\varphi)\, f_{\rm
flux}(v,\varphi)\, . \label{flux}
\end{equation}
Here $P_{\rm equi}$ denotes the equilibrium distribution
and $f_{\rm flux}$ describes deviations from it such that
it tends to 0 towards the continuum and to 1 towards the
well. The flux state, as a phase-space density, is
obtained as a stationary solution to the Fokker-Planck
equation (FPE) corresponding to (\ref{lange}).  In case of
vanishing non-Gaussian noise, $\eta(t)=0$, the latter one
reads ${\partial_t P(v,\varphi,t)}=L_0 P(v,\varphi,t)$
where
\begin{equation}
L_0=-v\partial_\varphi+\partial_v[V'(\varphi)/m+\gamma
v]+\gamma/(m\beta)
\partial^2_v \, .\label{barefpe}
\end{equation}
 Then, from $L_0 P_{\rm flux}=0$ one finds for moderate to strong
friction   the known expression \cite{revmodphys}
$\Gamma=(\omega_0\Omega/2\pi)\exp(-\beta V_b)$ with the
scaled Grote-Hynes frequency
$\Omega=-\gamma/(2\omega_b)+\sqrt{(\gamma/2\omega_b)^2+1}$.

For finite non-Gaussian noise the translation of
(\ref{lange}) into an equivalent FPE for the averaged
phase space distribution leads to only formal expressions
\cite{risken,hang}. Here, in accordance with the
experimental situation we proceed by assuming that (i)
non-Gaussian fluctuations are weak and (ii) sufficiently
fast compared to the bare dynamics of the system (detailed
conditions will be given below). Then, one considers
$V_{\rm eff}=V-\varphi \eta(t)$ as an effective
time-dependent potential so that for each realization of
the non-Gaussian random force a FPE for
 $P_\eta(v,\varphi,t)$ exists where $L_0$ is replaced by
 $L_0-\eta(t)/m\,
\partial_v$. To gain a FPE for the averaged distribution
$\langle P_\eta(v,\varphi,t)\rangle$, we switch to the
interaction picture $ P_\eta(t)=\exp[{L_0 t}]\, Q_\eta(t)$
with $\partial_v(t)=\exp(-L_0 t)\partial_v\exp(L_0 t)$. By
averaging the equation for $Q_\eta(t)$ over the
$\eta$-noise, the exact solution is expressed in terms of
the generating functional with the counting field $w(s)$
substituted by $(i/m)\partial_v(s)$, i.e.,
\begin{eqnarray}
\langle Q_\eta(t)\rangle&=&{\rm
e}^{-S[(i/m)\partial_v(s)]}\
Q(0)\nonumber\\
&=&{\cal T} \exp\left[\int_0^t\!\!\! ds \sum_{k\geq 2}
\frac{C_k(s)}{m^k}\right] {Q}(0)\,  .\label{qsolu}
\end{eqnarray}
Here, we have used the cumulant expansion of
$S[(i/m)\partial_v(s)]$ with $C_k(s)=\exp[-L_0 s]
\hat{C}_k(s)\exp[L_0 s]$, where the two lowest order
cumulant operators read
\begin{eqnarray}
\hat{C}_2(s)&=&\int_0^s du\, S_2(u)\, \partial_v\ \partial_v(-u)\nonumber\\
\hat{C}_3(s)&=&-\int_0^s du \int_0^{s-u}\!\!\!\!\!\!\!\!
du'\, S_3(u,u')
\partial_v\
\partial_v(-u)\ \partial_v(-u-u')\, .\nonumber
\end{eqnarray}
The above result reveals that the generator for the
averaged dynamics is directly given by the generating
functional of the non-Gaussian noise. Its cumulant
expansion now allows for a systematic approximation.

To do so, we take only the two leading order terms into
account in (\ref{qsolu}), which in turn leads to
$\partial_t \langle P_\eta\rangle=\left[
L_0+\hat{C}_2(t)/m^2+\hat{C}_3(t)/m^3\right]\, \langle
P_\eta\rangle$.
 While this
FPE  applies to weak non-Gaussian noise of arbitrary type,
following assumption (ii), the noise correlation functions
vanish on a sufficiently short time scale so that the
non-Markovian dynamics reduces effectively to a Markovian
one. This way, one arrives at a generalized FPE of the form
$\partial_t \langle P\rangle =L_{\rm eff} \langle
P\rangle$ with
\begin{equation}
L_{\rm eff}=
L_0+(c_2/m^2)\partial_v^2-(c_3/m^3)\partial_v^3\,
\label{leff}
\end{equation}
where
\[
c_2=\int_0^\infty du S_2(u)\ ,\  c_3=\int_0^\infty
du\int_0^\infty du' S_3(u,u')\, .\label{coeff}
\]
Note that this expression generalizes previous results
obtained in the overdamped limit \cite{hang} to weaker
friction. If the external noise were purely Gaussian,
i.e.\ $c_3=0$, the operator $L_{\rm eff}$ were exact.
Hence, what we really have to assume in deriving $L_{\rm
eff}$ are small higher than second order cumulants, while
$c_2$ may be arbitrarily large. Thus, one introduces an
effective temperature
\begin{equation}
T_{\rm eff}=T+c_2/(k_{\rm B} m\gamma)\,
 \label{teff}
\end{equation}
and incorporates the Gaussian components of the
non-Gaussian noise into a renormalized diffusion term
according to  $L_0(\beta)+(c_2/m^2)\partial_v^2 \to
L_0(\beta_{\rm eff})$. We note in passing that this
expression coincides with the one derived for resonant
activation \cite{resoact}. Experimentally, the heating due
to $c_2$ is substantial so that
 (\ref{teff}) is required to capture the actual temperature
 of the JJ.

With the generalized FPE at hand we now attack the rate
calculation. For this purpose, it is convenient to work
with dimensionless quantities $\tau=\omega_0 t$, $x=\varphi
\sqrt{\beta_{\rm eff} m\omega_0^2}$, $p=v\sqrt{\beta_{\rm
eff} m}$, $U=\beta_{\rm eff} V$, and
$\rho=\gamma/\omega_0$, $\bar{c}_3=c_3 (
\beta_{\rm eff}/m)^{3/2}$; then $x_b, U_b\gg 1$ and
$\bar{c}_3$ serves as a small parameter. We start with
 the equilibrium state (no flux) which determines the dominating
 exponential activation
factor in (\ref{flux}) and write $P_{\rm equi}\propto
P_{\beta_{\rm eff}} \exp(-\bar{c}_3 G)$ with
 the Boltzmann distribution $P_{\beta_{\rm eff}}$.
   Upon inserting this ansatz into $L_{\rm eff}P_{\rm
equi}=0$ we find  perturbatively a distribution of the form
 \begin{equation}
G(x,p)=\phi_0(x)+\sum_{n=1}^3\phi_n(x) p^n/n\, ,
\label{ffrom}
\end{equation}
where coordinate dependent functions can be expressed in
terms of only $\phi_2$ as
\begin{eqnarray}
\phi_1(x)&=& -\rho \phi_2(x)/U'(x)=-\phi_3(x) \nonumber\\
\phi_0(x)&=& 3x-3\rho\int_0^x\!\!\!\!\!\! dy
\phi_1(y)+\int_0^x\!\!\!\!\!\!dy U'(y)\phi_2(y)\, .
\label{solu2}
\end{eqnarray}
Note that the property $\phi_1=-\phi_3$ ensures that
$\langle p\rangle_{\rm equi}=0$ in leading order. While the
above expressions hold in general, the solution of
$\phi_2$ depends on the specific form of the potential $U$.
To better understand the deviations from the thermal
equilibrium we thus first study a harmonic potential
$U(x)=x^2/2$, which  approximates a metastable potential
around its well minimum. The result is $\phi_2(x)=-2
x/(1+2\rho^2)$ and the corresponding distribution $P_{\rm
equi}$ has a simple interpretation: it coincides with the
thermal distribution driven by the external force
$\eta(t)$ and averaged in the long time limit over the
non-Gaussian noise. Specifically, one has $P_{\rm
equi}(p,x)=\lim_{\tau\to\infty}\langle P_{\beta_{\rm eff}}
(p_\tau,x_\tau)\rangle$, where $p_\tau=p- \dot{f}(\tau)$
and $x_\tau=x- f(\tau)$ with $f(\tau)=\sqrt{\beta_{\rm
eff}/m\omega_0^2}\int_0^\infty d\tau'
\chi(\tau-\tau')\eta(\tau')$ and $\chi(\tau)$ the response
function for a damped harmonic oscillator. As a
consequence, $P_{\rm equi}(p,x)$  shows an asymmetry even
for a symmetric potential depending {\em linearly} on the
third cumulant $S_3$. In contrast, in a set-up where for
the $\eta$ noise a dissipation fluctuation theorem and
correspondingly a detailed balance relation applies, a
linear dependence on odd cumulants only occurs on a
transient time scale \cite{brosco}.

A generic metastable potential is of the form
$U(x)=(x^2/2)(1-2 x/3 x_b)$ describing particularly a
well-barrier segment of a tilted washboard potential of a
JJ. Then,
\begin{equation}
\phi_{2}(x)=-\frac{2x_b z(1-z)^{a}}{1+a}\,
_2F_1(a,1+a,2+a,z)\, \label{exactsolu}
\end{equation}
where $z=x/x_b$, $a=2\rho^2$ and $P_{\rm equi}$ follows
together with (\ref{solu2}). There is a little subtlety
here in that the solution (\ref{exactsolu}) applies for
$2\rho^2<1$ only outside a narrow range around $x_b$. The
global solution is then constructed by properly matching
the local solution around $x_b$ onto the latter one. Since
the vicinity around the top affects only the prefactor and
not the experimentally dominant exponential factor of the
rate, we will give further details elsewhere. This
exponential activation factor can now be inferred from
$P_{\rm equi}(p=0,x=x_b)$ and is  obtained as $
\Gamma\propto \exp[-U_b(1-g)] $
 with the third cumulant correction
 $g=-\bar{c}_3\phi_0(x_b)/U_b$ explicitly evaluated as
\begin{equation}
g(\rho)\approx  \frac{6\, \bar{c}_3}{5\rho^2+1}\, \frac{
U_b}{x_b} \, .\label{gfacapprox}
\end{equation}
The remaining prefactor to the rate is determined by
$f_{\rm flux}$ in (\ref{flux}) and the well population
$N_{\rm well}$. The latter one is easily calculated
 from $P_{\rm equi}(p,x)$ restricted to the
domain around the well bottom, while the former one is
derived from the local dynamics around the barrier top.
Eventually, the rate is found as
\begin{equation}
\Gamma=\frac{\omega_0\Omega}{2\pi}
\left(1-\sqrt{\frac{\pi}{2}}\,
\bar{c}_3\phi_2'(x_b)\frac{\Omega^2\rho\,\kappa}{9+2\rho^2}
\right){\rm e}^{-U_b(1-g)}\, ,\label{rateexp}
\end{equation}
where $
\kappa=2\Omega(9+4\rho^2)+\rho[8\rho^2+\sqrt{\rho\Omega}(7+2\rho^2)]+4$
and $\phi_2'(x_b)$ results for $2\rho^2\leq 1$  from the
matching procedure leading, e.g., for $2\rho^2\ll 1$ to
$\phi_2'(x_b)=1/\rho^2$.  The above expression is the main
finding of this Letter, namely, the thermal escape rate
out of a metastable well in presence of weak
($\bar{c}_3\ll 1/U_b$) and fast (correlation time
$\tau_c\ll 1/\omega_0$) non-Gaussian continuous noise. This
result indeed verifies that the detector transforms the
noise asymmetry into a rate asymmetry depending on the
sign of the third cumulant $\bar{c}_3$. For $\bar{c}_3>0
[<0]$ the noise distribution favors large positive
(negative) fluctuations so that  the barrier is
effectively reduced (enhanced) and the distribution in the
metastable potential develops a tail towards large
positive (negative) momenta and coordinates causing a rate
increase (decrease).

\begin{figure}
\vspace*{0.1cm} \epsfig{file=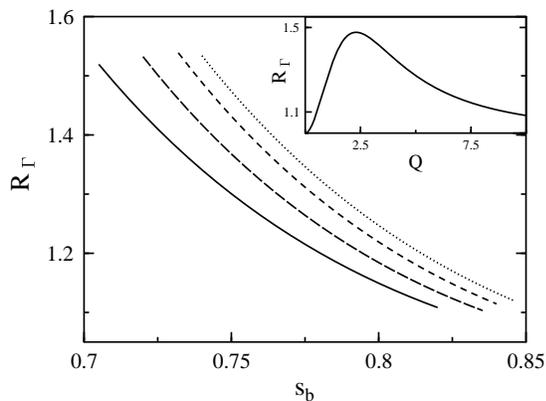, height=5.3cm}
\caption[]{\label{fig1} Rate asymmetry $R_\Gamma$
according to  (\ref{asymrate}) for
$\omega_0\hbar\beta=0.5$, $\beta E_J=200$, $Q=2.5$, and
various $s_m$ vs.\ the bias current $s_b$: $s_m=3$
(solid), $s_m=2.3$ (long-dashed), $s_m=1.7$
(short-dashed), $s_m=1.1$ (dotted); Inset shows $R_\Gamma$
for the same parameters and $s_b=0.75, s_m=1.1$ as a
function of $Q$.} \vspace*{-0.25cm}
\end{figure}

In the sequel, the general result (\ref{rateexp}) is
applied to a circuit, where a JJ acts as detector for the
current noise produced by a normal tunnel junction subject
to a voltage $V$ in the shot noise regime $e V\gg k_{\rm B}
T$. Further, $e V/\hbar \gg \omega_0$ guarantees that the
noise is much faster than the plasma frequency
$\omega_0=(\sqrt{2E_J E_C}/\hbar)(1-s_b^2)^{1/4}$
($s_b=I_b/I_c$) of the JJ with charging energy
$E_C=2e^2/C$ (capacitance $C$) and that its back-action
onto the tunnel junction is negligible. Then, the current
statistics is purely Poissonian: during a (scaled) time
interval $\tau_p$ an average number of $N$ charges passes
the conductor so that $s_m\equiv \langle I_m/I_c\rangle=N
e\omega_0/(\tau_p I_c)$ and
$S_3(\tau,\tau')=\delta(\tau+\tau')\delta(\tau)
(e\omega_0/I_c)^2\, s_m$.   The procedure to extract the
third cumulant is to measure switching rates with bias
currents $s_b$ (or equivalently mesoscopic currents $s_m$)
 flowing forward and backwards, respectively \cite{saclay}. This way,
one finds for the rate asymmetry
$R_\Gamma=\Gamma(|s_b|)/\Gamma(-|s_b|)$ as the dominant
contribution $R_\Gamma\approx \exp[2 U_b g(|s_b|)]$. In
the case considered here the result for this asymmetry
expressed in junction parameters reads
\begin{equation}
R_\Gamma=\exp\left[\frac{4\sqrt{2}\, \beta_{\rm eff}^3\,
E_C E_J^2\, Q^2}{3 (5+Q^2) }\frac{s_m\,
(1-|s_b|)^2}{\,\sqrt{1+|s_b|}}\right] \label{asymrate}
\end{equation}
with the quality factor $Q=1/\rho$ and the effective
temperature $T_{\rm eff}=T+Q \hbar\omega_0\, s_m/(4 k_{\rm
B} \sqrt{1-s_b^2})$. Note that in the exponent the ratio
between third cumulant and Gaussian noise basically
appears via $E_J^2 \beta_{\rm eff}^2 s_m$ so that
$R_\Gamma\to 1$ for increasing $s_m$.
 The above finding  not only allows to understand experimental data, but  may be
used to optimize the detection circuit as well. Namely, as
a function of $Q$, the asymmetry $R_\Gamma$ exhibits a
maximum for intermediate $Q$-values, but decreases towards
 higher (underdamped) and lower (overdamped) ones (Fig.~\ref{fig1}).
 This also reveals that  a rate calculation in
either of these limiting cases is not sufficient. For
typical experimental parameters
$s_m=2\omega_0\hbar\beta=1$, $\beta E_J=200$, and
$s_b=0.75$ one gains a maximal $R_\Gamma\approx 1.45$ at
$Q\approx 2.5$. Asymmetry ratios for various $s_m$, see
Fig.~\ref{fig1}, lie around this value in agreement with
recent experimental findings \cite{saclay}. Note that for
these parameters $T_{\rm eff}/T\approx 1.5$. Precise
numerical simulations confirm the result (\ref{asymrate})
and will be discussed elsewhere \cite{waintal}.
The general results (\ref{gfacapprox}), (\ref{rateexp})
determine  the rate asymmetry due to the third cumulant
 also for other mesoscopic conductors such as diffusive wires
or chaotic cavities. Since they apply for fast
fluctuations, they provide a tool  to analyse current
noise measurements in the interesting {\em high frequency}
regime \cite{buttiker2}.

To summarize, we have developed a formalism to describe the
switching process of a JJ  in presence of weak and fast
non-Gaussian fluctuations. From the corresponding rate
expression the rate asymmetry due to the third moment of
current noise has been obtained. Our results for the
generalized FPE (\ref{leff})  and the escape rate
(\ref{rateexp}) are applicable to other decay processes in
physics and chemistry,  where complex noise sources are
present.

Fruitful discussions with H. Pothier, B. Huard, D. Esteve,
N. Birge, J. Pekola, and H. Grabert are gratefully
acknowledged. JA is a Heisenberg fellow of the DFG.

\end{document}